\documentstyle[epic,eepic,epsfig,twocolumn,aps,pre]{revtex}
  \draft   
  \title{Shape Dynamics of Interfacial Front in Rotating Cylinders}
  \author{Gerald H. Ristow$^1$ and Masami Nakagawa$^2$}
  \address{$^1$Fachbereich Physik, Philipps-Universit\"at, Renthof 6, 
           35032 Marburg, Germany}
  \address{$^2$Division of Engineering, Colorado School of Mines, Golden, CO 
           80401, USA}
  \date{received 9. January 1998; revised 12. August 1998; received in final
  form \today}
\begin{document}
\maketitle
\begin{abstract}
The evolution of the interface propagation in a slowly rotating half-filled
horizontal cylinder is studied using MRI. Initially, the cylinder contains two
axially segregated bands of small and large particles with a sharp interface.
The process of the formation of the radial core is clearly captured, and the
shape and the velocity of the propagating front are calculated by assuming a
one-dimensional diffusion process along the rotation axis of the cylinder and
a separation of time scales associated with segregation in the radial and
axial directions. We found that the interfacial dynamics are best described
when a concentration dependent diffusion process is assumed.
\end{abstract}

\vspace*{-0.5cm}

\section{Introduction}
When granular materials are put in a horizontal rotating cylinder, most
particles undergo a solid body rotation except for a thin layer of particles
flowing near the free surface. With increasing rotation rate, particles first
exhibit intermittent avalanches, then continuous flow near the surface and
finally a ring structure due to centrifugal force~\cite{rajchenbach90}. In
industrial applications, rotating cylinders are primarily used for mixing
different components~\cite{bridgwater76}. But it is well known that particles
of different sizes or density show radial segregation on small time scales and
axial segregation on large time scales which compete against the mixing
process~\cite{bridgwater76,oyama39,donald62,nakagawa94,clement95,fabi97,%
choo97,hill97}.

One of the ways to investigate the three-dimensional particle motion is to
stop the rotating cylinder and take samples from different
locations~\cite{hogg66}.However, this inevitably involves a partial
destruction of the internal structure which might provide vital information.
Another shortfall of this approach is that it is time consuming to obtain a
global information about the particle motion. Recently, non-invasive {\em
magnetic resonance imaging} (MRI) has been used to interrogate the internal
structures of evolving granular assemblies~\cite{nakagawa93}. Since the
current MRI flow measurement is designed for a steady flow of particles and
recording a single image can take several minutes, the cylinder is also
stopped to capture a series of static images of the propagating front starting
with the same set of initial conditions. Nakagawa et al.~\cite{nakagawa93}
first applied MRI to study the dynamics of granular flow in a rotating
cylinder and recorded velocity and density profiles along and perpendicular to
the flowing layer of particles. Using a spin-tagging sequence, the flow and
diffusion in a vertically shaken container of poppy seeds was recently
measured and convection rolls found~\cite{ehrichs95}. Most recently, the shape
of the radially and axially segregated core were also investigated by
MRI~\cite{hill97,nakagawa97b}.

Not much is known yet about how this radially segregated core evolves.  In the
common segregation experiments, one starts with a well mixed initial state
containing small and large particles. The formation of a radial core of small
particles which extends over the whole length of the cylinder is already
visible after a few rotations~\cite{donald62}. This  radial core may become
unstable forming band patterns which is called {\em axial segregation}. This
process was best demonstrated by Hill et al.~\cite{hill97} using MRI. The final
number of bands, their positions and widths varied from experiment to
experiment but Nakagawa~\cite{nakagawa94} found the three band configuration to
be stable after an extended number of rotations. These bands may not be pure
and a radial core might still be present whereas Chicharro et al.~\cite{fabi97}
rotated two sizes of Ottawa sand for two weeks at 45 rpm and found a final
state of two {\em pure} bands each filling approximately half of the cylinder,
i.e.\ {\em no} radial core was found.  Recently, experimental evidences were
given for two new mechanisms of axial segregation, namely {\em traveling
waves}~\cite{choo97} and {\em avalanche mediated transport}~\cite{frette97}
which were only observed when non-spherical particles were used.

Since the dynamics of the axial segregation process depends on the homogeneity
of the initial mixture, it is desirable to start with a better  defined initial
packing. We propose to start with a fully axially segregated state of a binary
particle mixture and use MRI to investigate the three-dimensional segregation
front which is initially seen as a sharp flat interface. We study the
propagation velocities and compare them with results obtained from a simple
diffusive process where a {\em concentration dependent} diffusion coefficient
is used. The shape of the segregation front can be calculated approximately
analytically when a separation of time scales for the segregation in the 
radial direction and the diffusion in the axial direction is assumed.

\begin{figure}[htb]
  \begin{center}
    \setlength{\unitlength}{0.00083333cm}
\begin{picture}(10246,2781)(0,-10)

\put(2273,1533){\shade\circle{636}}
\put(2873,1158){\shade\circle{636}}
\put(3473,1533){\shade\circle{636}}
\put(398,1308){\shade\circle{636}}
\put(848,633){\shade\circle{636}}
\put(2798,333){\shade\circle{636}}
\put(4148,1608){\shade\circle{636}}
\put(4223,333){\shade\circle{636}}
\put(4748,1083){\shade\circle{636}}
\put(3998,933){\shade\circle{636}}
\put(3473,408){\shade\circle{636}}
\put(1748,933){\shade\circle{636}}
\put(2123,408){\shade\circle{636}}
\put(4898,408){\shade\circle{636}}
\put(5498,633){\shade\circle{636}}
\put(1223,1608){\shade\circle{636}}

\put(5348,1608){\ellipse{336}{336}}
\put(4898,1758){\ellipse{336}{336}}
\put(5573,1308){\ellipse{336}{336}}
\put(5948,1083){\ellipse{336}{336}}
\put(6173,558){\ellipse{336}{336}}
\put(6023,258){\ellipse{336}{336}}
\put(6548,258){\ellipse{336}{336}}
\put(6923,483){\ellipse{336}{336}}
\put(7298,258){\ellipse{336}{336}}
\put(7598,633){\ellipse{336}{336}}
\put(7823,258){\ellipse{336}{336}}
\put(8123,558){\ellipse{336}{336}}
\put(8348,258){\ellipse{336}{336}}
\put(8723,558){\ellipse{336}{336}}
\put(9098,333){\ellipse{336}{336}}
\put(9398,558){\ellipse{336}{336}}
\put(9773,483){\ellipse{336}{336}}
\put(6398,1008){\ellipse{336}{336}}
\put(6248,1458){\ellipse{336}{336}}
\put(5798,1608){\ellipse{336}{336}}
\put(6923,1083){\ellipse{336}{336}}
\put(6623,1458){\ellipse{336}{336}}
\put(6398,1833){\ellipse{336}{336}}
\put(6998,1683){\ellipse{336}{336}}
\put(7298,1308){\ellipse{336}{336}}
\put(7523,933){\ellipse{336}{336}}
\put(7748,1383){\ellipse{336}{336}}
\put(7448,1683){\ellipse{336}{336}}
\put(7973,1758){\ellipse{336}{336}}
\put(8198,1158){\ellipse{336}{336}}
\put(7898,933){\ellipse{336}{336}}
\put(8348,1533){\ellipse{336}{336}}
\put(8573,1083){\ellipse{336}{336}}
\put(9023,1008){\ellipse{336}{336}}
\put(8798,1458){\ellipse{336}{336}}
\put(9473,1008){\ellipse{336}{336}}

\thicklines
\put(923,1383){\ellipse{1800}{2700}}
\put(9323,1383){\ellipse{1800}{2700}}
\put(5193.109,784.630){\arc{1509.785}{0.0022}{1.6638}}
\path(4223,1983)(5948,783)
\path(923,2733)(9323,2733)
\path(923,33)(9323,33)
\path(1748,783)(98,1983)
\path(98,1983)(8498,1983)
\path(8498,1983)(10148,783)
\path(1748,783)(10148,783)

\end{picture}
  \end{center}
  \caption{A simplified sketch of the initial configuration: large particles 
	   are all in the left half of the cylinder and shown in gray. This
	  corresponds to the first side view image in Fig.~\ref{fig:
	  mri_start}.}
  \label{fig: setup}
\end{figure}
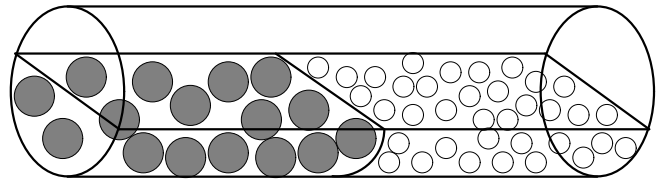
\section{Experimental Setup and Studies}
The initial state was prepared in such a way that the left half of the cylinder
contained only large particles having a diameter of 4 mm and the right half
contained only small particles having a diameter of 1 mm. It is similar to the
one used in~\cite{nakagawa97} and is sketched in Fig.~\ref{fig: setup} where the
larger particles are shown in gray. A 50-50 mixture by volume was used and the
total volume was chosen to give a roughly half-filled cylinder.

Axial migration of the core of small particles is a three dimensional process
and its dynamics should be monitored by a non-invasive method. The very
strength of MRI among other existing non-invasive techniques lies in its
ability of measuring dynamic flow properties such as velocities and  velocity
fluctuations. Measurement of true concentration still remains to be a
challenging problem because in granular flows concentration and particle
agitation, characterized by the granular temperature, are strongly coupled
which is predicted by kinetic theories. It is not trivial to separate causes
for the lesser concentration signals. Less challenging but equally important is
MRI's ability of probing a dense static granular system. In fact, the denser
the system of interest is, the higher the quality of images become. Since the 
MRI technique used is best suited for flows in a steady state, the cylinder was
stopped when the images were taken.

Our MRI imager/spectrometer (Nalorac Cryogenics Corp) has a 1.9 Tesla
superconducting magnet (Oxford) with a bore diameter of 31 cm. The useful
space inside the bore is a sphere of diameter of about 8 cm after the
insertion of the gradient coils and the rf probe. The acrylic cylinder with
10 cm and 7 cm in its length and diameter, respectively, was rotated at a
constant rate of 11.4 rpm by a dc servo motor (12FG) and a controller
(VXA-48-8-8) made by PMI inside the imaging apparatus. The material
properties, the drum geometry and its rotation speed were chosen such that
the drum was operating in the beginning of the continuous flow regime. Images
were taken after 0, 15, 30, 45, 60, 75, 90 and 600 seconds of rotation. The
particles used were spherical pharmaceutical pills of 1 and 4 mm diameter
which contain a liquid core of vitamin oil. Louge et al.~\cite{louge97}
conducted a detailed binary impact experiment using these particles and
estimated the normal coefficient of restitution to be about 0.9. In the
perpendicular direction, however, they have found that these liquid-filled
particles exhibit rolling contacts with negligible compliance.

\begin{figure}[htb]
  \hspace*{-0.04\textwidth}\psfig{file=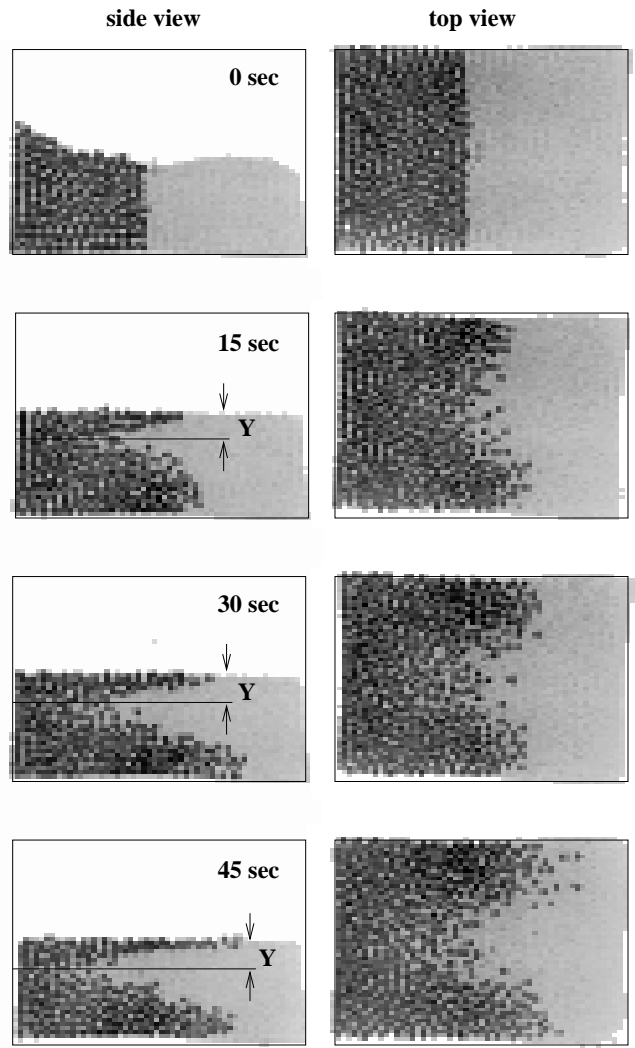,width=0.5\textwidth}
  \caption{Cross sectional views of the interface propagation for a mixture of 1 
	   and 4 mm vitamin pills in a half-filled rotating cylinder. The small 
	  particles cannot be resolved individually and the region they occupy 
	  is shown in light gray. }
  \label{fig: mri_start}
\end{figure}

The cylinder was put into the magnet for recording three-dimensional intensity
signals (64x64x64 points) which took around 40 minutes each. Due to the
limitations in the spatial resolution, the smaller particles cannot be resolved
individually and appear light gray in the images. Since the larger particles
contain more liquid they give a higher signal and show up in black or dark gray
in the images. Despite the fact that our images are three-dimensional, we have
decided to illustrate the dynamics of the conical shape of the migrating front
by showing two different cross sectional views in Fig.~\ref{fig: mri_start},
namely the side views on the left and the top views on the right. Initially (0
sec), the large and small particles were packed in the left and right half of
the cylinder, respectively, and the heaping of the large particles at the left end in the top
left image is due to an actual uneven initial packing. This quickly disappeared
and was flattened to the same level as in other parts of the cylinder once the
cylinder was in motion.

The side views were obtained by superimposing the 16 vertical planes in the
central region of the cylinder and taking the highest intensity value of all
planes. This averaging increased the contrast and we verified that the shape
of the front did not change significantly within these planes. For the top
view, 12 planes in the lower half of the cylinder were superimposed. By this
procedure, we can clearly identify the regions containing large particles
without loosing spatial information. The time evolution in Fig.~\ref{fig:
mri_start} shows that the cascading layer is almost entirely composed of large
particles whereas the small particles can {\em sink into} the surface layer
more easily and consequently their concentration in the layer will decrease.
The distance of the tip of the migrating front of small particles and the free
surface, denoted by Y in Fig.~\ref{fig: mri_start}, remains more or less
constant. This indicates that the small particles in the cascading layer are
responsible for the front advancement. Immediately after the cylinder starts
to rotate, there is a mixture of large and small particles flowing in the
cascading layer developed at the interface. Due to percolation mechanism, the
smaller particles travel under the free surface and move in the axial
direction driven by the concentration gradient. The smaller particles advance
more easily into the region occupied by larger particles since there are more
voids in the cascading region for the small particles to move into. We also
observed that the larger particles traveling on the surface always reached the
other end of the cylinder before the smaller particles did. Eventually, this
leads to an extended radially segregated core, shown in Fig.~\ref{fig:
mri_end} and also observed in other experiments~\cite{hill97}.
\begin{figure}[htb]
  \hspace*{-0.05\textwidth}\psfig{file=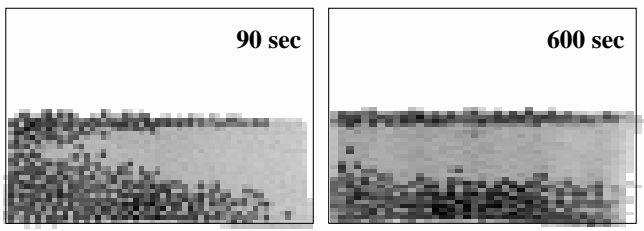,width=0.55\textwidth}
  \caption{Side view of the segregation front for later times at 90 s (left) 
	   and at 600 s (right) after start of rotation.}
  \label{fig: mri_end}
\end{figure}

\section{Radial Core Development}
Assuming random particle motion in the axial direction ($z$ axis), one component 
systems could be well described by a diffusion process according to Fick's 
Second Law~\cite{hogg66,cahn67}. The interface of a two component system can also 
be studied in this fashion and the diffusion equation reads
\begin{equation}
  \frac{\partial C(z,t)}{\partial t} = \frac{\partial}{\partial z} \left( D 
        \frac{\partial C(z,t)}{\partial z} \right)
  \label{eq: diff}
\end{equation}
where $C(z,t)$ and $D$ denote the relative concentration by volume of the 
smaller particles and the corresponding diffusion coefficient, respectively.
The initial condition for a cylinder with length $L$ are
\[
   C(z,0) = \left\{ \begin{array}{ll}
                     0, & \quad -\frac{L}{2} \le z < 0 \\
		     1, & \quad 0 < z \le \frac{L}{2}
                    \end{array} \right.
\]
whereas the boundary conditions read
\[ \left. \frac{\partial C}{\partial z} \right|_{z=-\frac{L}{2}} =
   \left. \frac{\partial C}{\partial z} \right|_{z=\frac{L}{2}} = 0
\]
which states that there is zero axial flux at the boundaries due to the end
caps.

For a constant diffusion coefficient, Eq.~(\ref{eq: diff}) can be solved
analytically for the specified initial and boundary  conditions and the
solution reads
\begin{eqnarray}
  C(z,t) = \frac{1}{2} + \frac{2}{\pi} \sum_{k=1}^\infty \frac{1}{2k-1}
  &\exp&\left\{-\frac{(2k-1)^2\pi^2 D t}{L^2}\right\}\times\nonumber\\ 
  &\sin&\left\{\frac{(2k-1)\pi z}{L}\right\} \ .
  \label{eq: profile}
\end{eqnarray}
Hogg et al.~\cite{hogg66} solved a similar diffusion problem for a single
component system by replacing the time derivative with the derivative with
respect to the number of rotations thus assuming that the dynamic effects on
the diffusion process was not significant within the range of the rotation
rates studied. Nakagawa et al.~\cite{nakagawa97} investigated a mixing process
of 50--50 mixtures of liquid filled pharmaceutical pills of different sizes.
After different numbers of revolutions they inserted 12 dividers along the
rotational axis and measured the fraction of small particles in each of the 13
segments. They obtained one-dimensional concentration profiles similar to the
theoretical prediction given by Eq.~(\ref{eq: profile}). However, for longer
rotation times, their profiles showed a slight asymmetry with respect to the
position of the initial front. It should be kept in mind, however, that the
final configuration in both our MRI study and Ref.~\cite{nakagawa97} was a
radial segregated core of small particles and not a mixture as in
Refs.~\cite{hogg66,cahn67}."

It was found experimentally~\cite{bridgwater76,clement95} and
numerically~\cite{dury97} that a nearly complete radial segregation can be
achieved after only a few rotations. In the above, successive MRI pictures
were separated by 15 seconds which corresponds to roughly 3 full rotations and
each particle had a chance to participate in approximately 6 avalanches. In
order to calculate analytically the exact shape of the interface of small
particles migrating into the region initially occupied by large particles, it
is assumed that the concentration change in each section is slow enough that
the fully radially segregated core is already completely developed. This
approximation states a direct relationship between concentration $C(z,t)$ and
width and position of the region occupied by small particles which will be
given now.

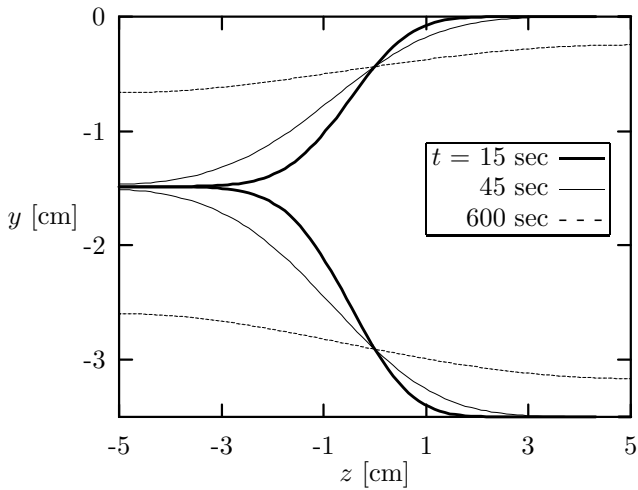
\begin{figure}[htb]
  \begin{center}
\setlength{\unitlength}{0.240900pt}
\begin{picture}(1087,765)(0,0)
\thicklines \path(220,203)(240,203)
\thicklines \path(1023,203)(1003,203)
\put(198,203){\makebox(0,0)[r]{-3}}
\thicklines \path(220,383)(240,383)
\thicklines \path(1023,383)(1003,383)
\put(198,383){\makebox(0,0)[r]{-2}}
\thicklines \path(220,562)(240,562)
\thicklines \path(1023,562)(1003,562)
\put(198,562){\makebox(0,0)[r]{-1}}
\thicklines \path(220,742)(240,742)
\thicklines \path(1023,742)(1003,742)
\put(198,742){\makebox(0,0)[r]{0}}
\thicklines \path(220,113)(220,133)
\thicklines \path(220,742)(220,722)
\put(220,68){\makebox(0,0){-5}}
\thicklines \path(381,113)(381,133)
\thicklines \path(381,742)(381,722)
\put(381,68){\makebox(0,0){-3}}
\thicklines \path(541,113)(541,133)
\thicklines \path(541,742)(541,722)
\put(541,68){\makebox(0,0){-1}}
\thicklines \path(702,113)(702,133)
\thicklines \path(702,742)(702,722)
\put(702,68){\makebox(0,0){1}}
\thicklines \path(862,113)(862,133)
\thicklines \path(862,742)(862,722)
\put(862,68){\makebox(0,0){3}}
\thicklines \path(1023,113)(1023,133)
\thicklines \path(1023,742)(1023,722)
\put(1023,68){\makebox(0,0){5}}
\thicklines \path(220,113)(1023,113)(1023,742)(220,742)(220,113)
\put(45,427){\makebox(0,0)[l]{\shortstack{$y$ [cm]}}}
\put(621,23){\makebox(0,0){$z$ [cm]}}
%
%
\Thicklines \path(220,475)(220,475)(228,475)(236,475)(244,475)(252,475)(261,475)(269,475)(277,475)(285,475)(293,475)(301,475)(309,475)(317,475)(325,475)(334,475)(342,474)(350,474)(358,474)(366,473)(374,473)(382,472)(390,471)(398,470)(407,469)(415,467)(423,465)(431,463)(439,460)(447,457)(455,453)(463,448)(471,443)(480,437)(488,430)(496,422)(504,413)(512,403)(520,393)(528,381)(536,369)(544,356)(553,343)(561,328)(569,314)(577,299)(585,284)(593,269)(601,254)(609,240)(617,226)
\Thicklines \path(617,226)(626,212)(634,200)(642,188)(650,177)(658,168)(666,159)(674,151)(682,144)(690,138)(699,133)(707,129)(715,125)(723,122)(731,120)(739,118)(747,117)(755,116)(763,115)(772,114)(780,114)(788,114)(796,113)(804,113)(812,113)(820,113)(828,113)(836,113)(845,113)(853,113)(861,113)(869,113)(877,113)(885,113)(893,113)(901,113)(909,113)(918,113)(926,113)(934,113)(942,113)(950,113)(958,113)(966,113)(968,113)
\Thicklines \path(1004,113)(1007,113)(1015,113)(1023,113)
\Thicklines \path(220,475)(220,475)(228,475)(236,475)(244,475)(252,475)(261,475)(269,475)(277,475)(285,475)(293,475)(301,475)(309,475)(317,475)(325,475)(334,475)(342,476)(350,476)(358,476)(366,476)(374,477)(382,477)(390,478)(398,479)(407,479)(415,481)(423,482)(431,484)(439,486)(447,488)(455,491)(463,495)(471,499)(480,503)(488,509)(496,514)(504,521)(512,528)(520,536)(528,544)(536,553)(544,563)(553,573)(561,583)(569,594)(577,605)(585,616)(593,627)(601,638)(609,649)(617,659)
\Thicklines \path(617,659)(626,669)(634,678)(642,687)(650,695)(658,702)(666,708)(674,714)(682,719)(690,723)(699,727)(707,730)(715,733)(723,735)(731,737)(739,738)(747,739)(755,740)(763,740)(772,741)(780,741)(788,741)(796,742)(804,742)(812,742)(820,742)(828,742)(836,742)(845,742)(853,742)(861,742)(869,742)(877,742)(885,742)(893,742)(901,742)(909,742)(918,742)(926,742)(934,742)(942,742)(950,742)(958,742)(966,742)(968,742)
\Thicklines \path(1004,742)(1007,742)(1015,742)(1023,742)
%
%
\thinlines \path(220,470)(220,470)(228,470)(236,470)(244,469)(252,468)(261,468)(269,467)(277,465)(285,464)(293,463)(301,461)(309,459)(317,457)(325,455)(334,453)(342,450)(350,447)(358,444)(366,441)(374,437)(382,433)(390,429)(398,425)(407,420)(415,415)(423,410)(431,404)(439,398)(447,392)(455,385)(463,379)(471,371)(480,364)(488,357)(496,349)(504,341)(512,333)(520,324)(528,316)(536,307)(544,299)(553,290)(561,281)(569,273)(577,264)(585,256)(593,247)(601,239)(609,231)(617,223)
\thinlines \path(617,223)(626,215)(634,208)(642,201)(650,194)(658,187)(666,181)(674,175)(682,169)(690,164)(699,159)(707,154)(715,150)(723,146)(731,142)(739,139)(747,136)(755,133)(763,130)(772,128)(780,126)(788,124)(796,123)(804,121)(812,120)(820,119)(828,118)(836,117)(845,116)(853,116)(861,115)(869,115)(877,115)(885,114)(893,114)(901,114)(909,114)(918,114)(926,113)(934,113)(942,113)(950,113)(958,113)(966,113)(974,113)(982,113)(991,113)(999,113)(1007,113)(1015,113)(1023,113)
\thinlines \path(220,479)(220,479)(228,479)(236,479)(244,479)(252,480)(261,481)(269,481)(277,482)(285,483)(293,484)(301,485)(309,487)(317,488)(325,490)(334,491)(342,493)(350,495)(358,498)(366,500)(374,503)(382,506)(390,509)(398,512)(407,516)(415,519)(423,523)(431,527)(439,532)(447,536)(455,541)(463,546)(471,551)(480,557)(488,562)(496,568)(504,574)(512,580)(520,586)(528,592)(536,599)(544,605)(553,611)(561,618)(569,624)(577,631)(585,637)(593,643)(601,649)(609,655)(617,661)
\thinlines \path(617,661)(626,667)(634,672)(642,677)(650,683)(658,687)(666,692)(674,696)(682,701)(690,705)(699,708)(707,712)(715,715)(723,718)(731,720)(739,723)(747,725)(755,727)(763,729)(772,731)(780,732)(788,734)(796,735)(804,736)(812,737)(820,738)(828,738)(836,739)(845,739)(853,740)(861,740)(869,741)(877,741)(885,741)(893,741)(901,741)(909,741)(918,742)(926,742)(934,742)(942,742)(950,742)(958,742)(966,742)(974,742)(982,742)(991,742)(999,742)(1007,742)(1015,742)(1023,742)
%
%
\thinlines \dashline[0]{15}(220,275)(220,275)(228,275)(236,275)(244,275)(252,275)(261,274)(269,274)(277,273)(285,273)(293,272)(301,272)(309,271)(317,271)(325,270)(334,269)(342,268)(350,267)(358,266)(366,265)(374,264)(382,263)(390,262)(398,261)(407,259)(415,258)(423,257)(431,255)(439,254)(447,253)(455,251)(463,250)(471,248)(480,247)(488,245)(496,244)(504,242)(512,241)(520,239)(528,238)(536,236)(544,234)(553,233)(561,231)(569,229)(577,228)(585,226)(593,225)(601,223)(609,221)(617,220)
\thinlines \dashline[0]{15}(617,220)(626,218)(634,217)(642,215)(650,214)(658,212)(666,211)(674,209)(682,208)(690,206)(699,205)(707,203)(715,202)(723,200)(731,199)(739,198)(747,196)(755,195)(763,194)(772,193)(780,192)(788,190)(796,189)(804,188)(812,187)(820,186)(828,185)(836,184)(845,183)(853,182)(861,182)(869,181)(877,180)(885,179)(893,179)(901,178)(909,177)(918,177)(926,176)(934,176)(942,175)(950,175)(958,175)(966,174)(974,174)(982,174)(991,174)(999,173)(1007,173)(1015,173)(1023,173)
\thinlines \dashline[0]{15}(220,623)(220,623)(228,623)(236,623)(244,623)(252,623)(261,623)(269,623)(277,624)(285,624)(293,624)(301,625)(309,625)(317,626)(325,626)(334,627)(342,628)(350,628)(358,629)(366,630)(374,631)(382,631)(390,632)(398,633)(407,634)(415,635)(423,636)(431,637)(439,638)(447,639)(455,640)(463,641)(471,642)(480,643)(488,644)(496,646)(504,647)(512,648)(520,649)(528,650)(536,651)(544,653)(553,654)(561,655)(569,656)(577,657)(585,659)(593,660)(601,661)(609,662)(617,663)
\thinlines \dashline[0]{15}(617,663)(626,664)(634,666)(642,667)(650,668)(658,669)(666,670)(674,671)(682,672)(690,673)(699,674)(707,675)(715,677)(723,678)(731,679)(739,679)(747,680)(755,681)(763,682)(772,683)(780,684)(788,685)(796,686)(804,687)(812,687)(820,688)(828,689)(836,689)(845,690)(853,691)(861,691)(869,692)(877,693)(885,693)(893,694)(901,694)(909,695)(918,695)(926,695)(934,696)(942,696)(950,696)(958,697)(966,697)(974,697)(982,697)(991,697)(999,697)(1007,697)(1015,698)(1023,698)
%
%
\put(702,460){\begin{tabular}{|rl|} \hline
		$t$~=~15 sec & \rule[0.5ex]{0.6cm}{0.6pt} \\
		      45 sec & \rule[0.5ex]{0.6cm}{0.2pt} \\
		     600 sec & - - - \\ \hline
		    \end{tabular}}

\end{picture}
  \end{center}
  \caption{Theoretical prediction of the interface shape for three different 
           times after start of rotation using the same constant diffusion 
	   coefficient $D=0.02$ cm$^2$/s.}
  \label{fig: theory_front}
\end{figure}

For a half-filled cylinder, the center of mass of the core of small particles
lies $\frac{4 R}{3 \pi}$ below the free surface. For lower concentrations, we
assume that all small particles occupy a region in the shape of a half circle,
again centered at the same point as the full half circle. The radius $r$ of
this half circle is related to the concentration and the cylinder radius by 
\[ C = \frac{r^2}{R^2}\ \Rightarrow\ r = R \sqrt{C} \ . \] The analytically
obtained interfacial shape dynamics of the region occupied by small particles
is shown in Fig.~\ref{fig: theory_front} for three different times after the
start of rotation. From the cross sectional views of Fig.~\ref{fig:
mri_start}, we estimated that the front of small particles reached the left
cylinder boundary around $t=45$ s which gave a value of $D=0.02$ cm$^2$/s for
the diffusion coefficient in Eq.~(\ref{eq: diff}). Mixing experiments using
6.4 mm Lucite beads gave diffusion coefficients around 0.1 cm$^2$/s and the
higher value is due to the larger particle diameter~\cite{cahn67}. All shapes
agree well with the MRI data considering the level of approximations made and
the dashed line for $t=600$ s predicts an already nearly fully radially
segregated core of small particles which corresponds well to the right picture
of Fig.~\ref{fig: mri_end}. In order to derive an improved model, however, we
need to know the exact shape of the region occupied by small particles and the
exact position of the center of mass of this region.

The proposed model with a constant diffusion coefficient $D$ in Eq.~(\ref{eq:
diff}) always leads to symmetric concentration profiles, which can be seen
from Eq.~(\ref{eq: profile}). However, the experimental data presented in
Ref.~\cite{nakagawa97} indicate that even though this is a good approximation
for small numbers of rotations, clear deviations from symmetric profiles are 
detected for large numbers of rotations. Since particle concentrations vary in
space and time and might have different particle packings due to the radial
segregation process, the mobility of the particles is also affected. The
proposed model enables us to test the effects of a {\em concentration
dependent} diffusion coefficient on the particle motion, especially the shape
dynamics. We will investigate two linear dependencies of the form
\begin{equation}
  D_{\pm}(C) = D_0 \pm D_0(C-0.5)
  \label{eq: diff_c}
\end{equation}
which fulfill
  \[ \langle D \rangle \equiv \int_0^1 D_{\pm}(C)\, dC = D_0 \]
and solve Eq.~(\ref{eq: diff}) numerically by using a standard
finite-difference procedure.

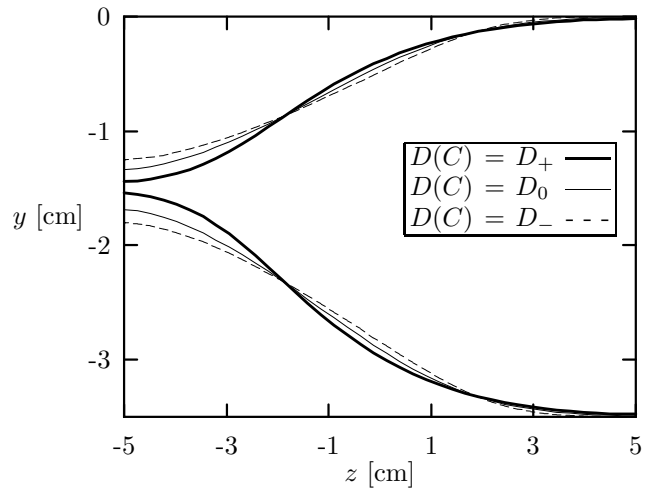
\begin{figure}[htb]
  \begin{center}
\setlength{\unitlength}{0.240900pt}
\begin{picture}(1087,765)(0,0)
\thicklines \path(220,203)(240,203)
\thicklines \path(1023,203)(1003,203)
\put(198,203){\makebox(0,0)[r]{-3}}
\thicklines \path(220,383)(240,383)
\thicklines \path(1023,383)(1003,383)
\put(198,383){\makebox(0,0)[r]{-2}}
\thicklines \path(220,562)(240,562)
\thicklines \path(1023,562)(1003,562)
\put(198,562){\makebox(0,0)[r]{-1}}
\thicklines \path(220,742)(240,742)
\thicklines \path(1023,742)(1003,742)
\put(198,742){\makebox(0,0)[r]{0}}
\thicklines \path(220,113)(220,133)
\thicklines \path(220,742)(220,722)
\put(220,68){\makebox(0,0){-5}}
\thicklines \path(381,113)(381,133)
\thicklines \path(381,742)(381,722)
\put(381,68){\makebox(0,0){-3}}
\thicklines \path(541,113)(541,133)
\thicklines \path(541,742)(541,722)
\put(541,68){\makebox(0,0){-1}}
\thicklines \path(702,113)(702,133)
\thicklines \path(702,742)(702,722)
\put(702,68){\makebox(0,0){1}}
\thicklines \path(862,113)(862,133)
\thicklines \path(862,742)(862,722)
\put(862,68){\makebox(0,0){3}}
\thicklines \path(1023,113)(1023,133)
\thicklines \path(1023,742)(1023,722)
\put(1023,68){\makebox(0,0){5}}
\thicklines \path(220,113)(1023,113)(1023,742)(220,742)(220,113)
\put(45,427){\makebox(0,0)[l]{\shortstack{$y$ [cm]}}}
\put(621,23){\makebox(0,0){$z$ [cm]}}
\Thicklines \path(220,483)(220,483)(241,484)(261,487)(282,491)(302,496)(323,502)(344,510)(364,519)(385,530)(405,542)(426,555)(446,569)(467,583)(488,597)(508,610)(529,623)(549,636)(570,647)(591,658)(611,668)(632,676)(652,685)(673,692)(694,699)(714,704)(735,710)(755,714)(776,719)(797,722)(817,725)(838,728)(858,730)(879,732)(899,734)(920,735)(941,737)(961,737)(982,738)(1002,738)(1023,739)
\Thicklines \path(220,465)(220,465)(241,463)(261,459)(282,454)(302,447)(323,438)(344,428)(364,415)(385,401)(405,384)(426,367)(446,348)(467,329)(488,310)(508,292)(529,274)(549,257)(570,242)(591,227)(611,214)(632,202)(652,191)(673,181)(694,172)(714,164)(735,157)(755,150)(776,145)(797,140)(817,136)(838,132)(858,129)(879,126)(899,124)(920,122)(941,120)(961,119)(982,118)(1002,118)(1023,118)
\thinlines \path(220,502)(220,502)(241,503)(261,506)(282,511)(302,516)(323,522)(344,529)(364,537)(385,545)(405,554)(426,564)(446,573)(467,584)(488,595)(508,605)(529,616)(549,627)(570,638)(591,649)(611,659)(632,669)(652,678)(673,686)(694,694)(714,702)(735,708)(755,714)(776,719)(797,723)(817,727)(838,730)(858,733)(879,735)(899,736)(920,738)(941,739)(961,740)(982,740)(1002,740)(1023,741)
\thinlines \path(220,439)(220,439)(241,437)(261,433)(282,427)(302,420)(323,411)(344,402)(364,391)(385,380)(405,368)(426,355)(446,342)(467,327)(488,313)(508,298)(529,283)(549,269)(570,254)(591,240)(611,226)(632,213)(652,200)(673,188)(694,178)(714,168)(735,159)(755,151)(776,144)(797,139)(817,133)(838,129)(858,126)(879,123)(899,120)(920,119)(941,117)(961,116)(982,116)(1002,115)(1023,115)
\thinlines \dashline[0]{15}(220,517)(220,517)(241,519)(261,521)(282,524)(302,529)(323,534)(344,540)(364,546)(385,553)(405,560)(426,568)(446,576)(467,584)(488,593)(508,602)(529,612)(549,621)(570,631)(591,640)(611,650)(632,660)(652,670)(673,679)(694,688)(714,697)(735,705)(755,713)(776,719)(797,725)(817,730)(838,733)(858,736)(879,738)(899,740)(920,741)(941,741)(961,742)(982,742)(1002,742)(1023,742)
\thinlines \dashline[0]{15}(220,418)(220,418)(241,416)(261,413)(282,408)(302,402)(323,395)(344,388)(364,379)(385,370)(405,360)(426,349)(446,338)(467,327)(488,315)(508,303)(529,290)(549,277)(570,264)(591,251)(611,237)(632,224)(652,211)(673,198)(694,186)(714,174)(735,163)(755,153)(776,144)(797,136)(817,130)(838,125)(858,121)(879,118)(899,116)(920,115)(941,114)(961,114)(982,113)(1002,113)(1023,113)
%
%
\put(660,460){\begin{tabular}{|rcll|} \hline
		$D(C)$ &=& $D_{+}$ & \rule[0.5ex]{0.6cm}{0.6pt} \\
		$D(C)$ &=& $D_0$   & \rule[0.5ex]{0.6cm}{0.2pt} \\
		$D(C)$ &=& $D_{-}$ & - - - \\ \hline
                    \end{tabular}}

\end{picture}
  \end{center}
  \caption{Theoretical prediction of the interface shape at $t=90$~s for three 
           different concentration dependencies of the diffusion coefficient 
	   $D$.}
  \label{fig: theory_front2}
\end{figure}

This linear concentration dependence has a significant effect on the shape
dynamics which is shown in Fig.~\ref{fig: theory_front2} by comparing both laws
with the shape given by a constant diffusion coefficient at $t=90$ s. The
effect is most pronounced in the left part of the cylinder, i.e.\ in the region
the small particles propagate into. When $D$ increases (decreases) with small
particle concentration $C$, the diameter of the segregated core close to the
left cylinder wall is larger (smaller). Unfortunately, we are not able to
judge, which of the investigated three concentration dependencies of $D$ gives
the best agreement with the experimental data. The spatial resolution is not
high enough to reconstruct the full shape which would be necessary in order to
distinguish a dependence of the form $D_{\pm}(C)$ from a constant diffusion
coefficient with a larger (for $D_{-}$) or smaller (for $D_{+}$) value.
However, we will present in the next section a more suitable method to do
so. 

\begin{figure}[htb]
  \begin{center}
\setlength{\unitlength}{0.240900pt}
\begin{picture}(1087,900)(50,0)
\footnotesize
\thicklines \path(177,249)(197,249)
\thicklines \path(1058,249)(1038,249)
\put(155,249){\makebox(0,0)[r]{5}}
\thicklines \path(177,388)(197,388)
\thicklines \path(1058,388)(1038,388)
\put(155,388){\makebox(0,0)[r]{6}}
\thicklines \path(177,527)(197,527)
\thicklines \path(1058,527)(1038,527)
\put(155,527){\makebox(0,0)[r]{7}}
\thicklines \path(177,666)(197,666)
\thicklines \path(1058,666)(1038,666)
\put(155,666){\makebox(0,0)[r]{8}}
\thicklines \path(177,806)(197,806)
\thicklines \path(1058,806)(1038,806)
\put(155,806){\makebox(0,0)[r]{9}}
\thicklines \path(177,945)(197,945)
\thicklines \path(1058,945)(1038,945)
\put(155,945){\makebox(0,0)[r]{10}}
\thicklines \path(177,179)(177,199)
\thicklines \path(177,945)(177,925)
\put(177,134){\makebox(0,0){0}}
\thicklines \path(353,179)(353,199)
\thicklines \path(353,945)(353,925)
\put(353,134){\makebox(0,0){20}}
\thicklines \path(529,179)(529,199)
\thicklines \path(529,945)(529,925)
\put(529,134){\makebox(0,0){40}}
\thicklines \path(706,179)(706,199)
\thicklines \path(706,945)(706,925)
\put(706,134){\makebox(0,0){60}}
\thicklines \path(882,179)(882,199)
\thicklines \path(882,945)(882,925)
\put(882,134){\makebox(0,0){80}}
\thicklines \path(1058,179)(1058,199)
\thicklines \path(1058,945)(1058,925)
\put(1058,134){\makebox(0,0){100}}
\thicklines \path(177,179)(1058,179)(1058,945)(177,945)(177,179)
\put(45,562){\makebox(0,0)[l]{\shortstack{$z$ [cm]}}}
\put(617,67){\makebox(0,0){$t$ [s]}}
\put(860,388){\makebox(0,0)[r]{MRI-data}}
\thinlines \path(177,197)(177,253)
\thinlines \path(167,197)(187,197)
\thinlines \path(167,253)(187,253)
\thinlines \path(309,509)(309,565)
\thinlines \path(299,509)(319,509)
\thinlines \path(299,565)(319,565)
\thinlines \path(441,605)(441,661)
\thinlines \path(431,605)(451,605)
\thinlines \path(431,661)(451,661)
\thinlines \path(573,653)(573,708)
\thinlines \path(563,653)(583,653)
\thinlines \path(563,708)(583,708)
\thinlines \path(706,701)(706,757)
\thinlines \path(696,701)(716,701)
\thinlines \path(696,757)(716,757)
\thinlines \path(838,749)(838,804)
\thinlines \path(828,749)(848,749)
\thinlines \path(828,804)(848,804)
\thinlines \path(970,797)(970,853)
\thinlines \path(960,797)(980,797)
\thinlines \path(960,853)(980,853)
\put(177,225){\circle*{12}}
\put(309,537){\circle*{12}}
\put(441,633){\circle*{12}}
\put(573,680){\circle*{12}}
\put(706,729){\circle*{12}}
\put(838,776){\circle*{12}}
\put(970,825){\circle*{12}}
\put(936,388){\circle*{12}}

\put(860,343){\makebox(0,0)[r]{$D=$ const.}}
\dashline[0]{15}(882,343)(990,343)
\dashline[0]{15}(177,249)(177,249)(191,327)(205,360)(221,388)(265,444)(308,488)(310,490)(352,526)(396,558)(440,587)(442,589)(575,666)(708,731)(840,788)(973,848)

\put(860,298){\makebox(0,0)[r]{$D=D_{-}$}}
\path(882,298)(990,298)
\path(177,249)(177,249)(183,309)(190,345)(196,361)(203,371)(230,410)(256,444)(283,475)(309,502)(336,521)(367,547)(441,600)(494,630)(558,666)(573,676)(706,741)(838,796)(970,849)

\path(670,420)(1020,420)(1020,260)(670,260)(670,420)

\end{picture}
  \end{center}
  \caption{Propagation of large particles along rotation axis, comparison of
	   experimental data ($\bullet$) and from diffusion equation using a 
	   constant ($\cdot\cdot\cdot$) or a concentration dependent (---) 
	   diffusion coefficient $D$.}
  \label{fig: mri_front}
\end{figure}
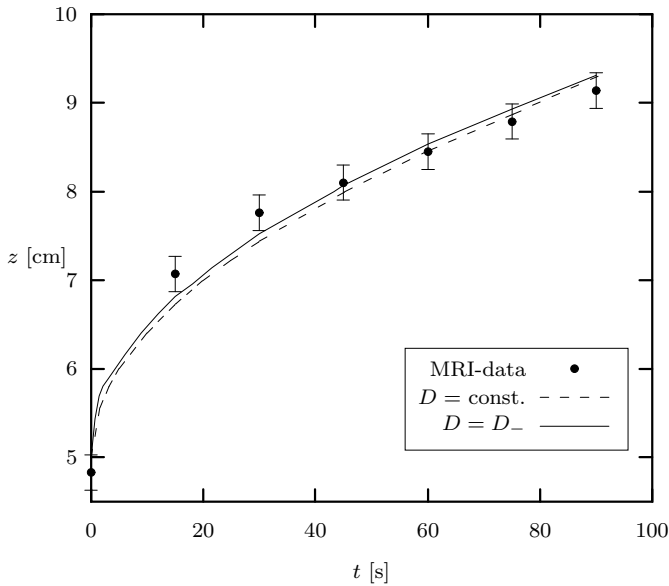

Since the larger particles give a higher MRI signal, the propagation of these 
particles into the region initially filled with small particles can be studied
by recording intensity values that correspond to large particles. These values
were extracted from the side views of Figs.~\ref{fig: mri_start} and~\ref{fig:
mri_end} and are shown in Fig.~\ref{fig: mri_front} as filled circles with
corresponding error bars. Also shown as dashed line is the theoretical curve
according to Eq.~(\ref{eq: profile}) using the same constant diffusion
coefficient of $D=0.02$ cm$^2$/s as in Fig.~\ref{fig: theory_front}. The data
shows that the front advancement is underestimated when a constant diffusion
coefficient is used. The particle dynamics are very different for particles in
the fluidized surface layer and for particles following the solid body
rotation. The fluidized surface layer has a thickness of a few large particles
for the range of rotation speeds used here. For a low concentration of small
particles, these can be trapped in a region well below this layer and  they
will show a low mobility relative to one another due to percolation. For
higher concentrations, the segregated core is large enough so that some of the
small particles in the core are mixed with the large particles in the
fluidized surface layer. Therefore, not all small particles can remain in the
segregated core but some will be recirculated in the flowing layer. The
mobility of particles in this layer is much higher than for particles in the
segregated core which makes us believe that the diffusion coefficient is
concentration dependent.

The exact functional dependence can only be inferred from a detailed
understanding of the microscopic particle motion of the two species, e.g.\ by
looking at the asymmetry of the concentration profile or by calculating the
diffusion coefficient microscopically. Unfortunately, neither the experimental
nor the numerical results available are accurate and detailed enough to
address this question fully. However, it is already very instructive to
consider the two linear dependencies proposed in Eq.~(\ref{eq: diff_c}) and we
show in Fig.~\ref{fig: mri_front} as solid line the numerical result for
$D(C)=D_{-}(C)$. The result for $D(C)=D_{+}(C)$ will lie between the solid and
dashed lines and in either case, we get a better agreement with the
experimental result when a constant diffusion coefficient $D(C)=D_0$ is used.

\section{Conclusions}
Using MRI measurements, we investigated the shape dynamics and interface
propagation in a slowly rotating half-filled horizontal cylinder. By assuming
that the radial segregation and the axial migration process occur on different
time scales, we showed that the propagating front can be well described by a
one-dimensional diffusion process.  The experimentally observed concentration
profiles are slightly asymmetrical~\cite{nakagawa97}, however, this can be 
understood by using a concentration dependent diffusion coefficient. We
demonstrated the effect when a linear dependence of the diffusion coefficient
on the particle concentration is used in the analytic description. In this
case, we found a better agreement with the experimental results as compared to
the case when the diffusion coefficient is constant. This implies that the
shape dynamics of the interface propagation can be described more accurately
when a concentration dependent diffusion coefficient is considered. However, in
order to determine the correct concentration dependence of the diffusion
coefficient, one must compare the shape of the whole concentration profile or
one has to calculate the diffusion coefficient microscopically from the
particle movements.

\section*{Acknowledgments}
We would like to thank everyone at The New Mexico Resonance (formerly a part
of The Lovelace Institutes) in Albuquerque for their kind hospitality during
our stay and especially acknowledge invaluable help with the experiments by
S.A. Altobelli. GR is supported by the Deutsche Forschungsgemeinschaft through
Ri 826/1-2. MN is supported in part by NASA through contract number NAG3-1970.

\vspace*{-0.5cm}


\end{document}